\documentclass{article}

\usepackage[english]{babel}
\usepackage{authblk}
\usepackage{listings}
\usepackage[letterpaper,top=2cm,bottom=2cm,left=3cm,right=3cm,marginparwidth=1.75cm]{geometry}
\usepackage{subcaption}

\usepackage{color}
\definecolor{codegreen}{rgb}{0,0.6,0}
\definecolor{codegray}{rgb}{0.5,0.5,0.5}
\definecolor{codepurple}{rgb}{0.58,0,0.82}
\definecolor{backcolour}{rgb}{0.95,0.95,0.92}

\lstdefinestyle{mystyle}{
    backgroundcolor=\color{backcolour},   
    commentstyle=\color{codegreen},
    keywordstyle=\color{magenta},
    numberstyle=\tiny\color{codegray},
    stringstyle=\color{codepurple},
    basicstyle=\ttfamily\footnotesize,
    breakatwhitespace=false,         
    breaklines=true,                 
    captionpos=b,                    
    keepspaces=true,                 
    numbers=left,                    
    numbersep=5pt,                  
    showspaces=false,                
    showstringspaces=false,
    showtabs=false,                  
    tabsize=2
}

\usepackage{algorithm}
\usepackage{algpseudocode}
\usepackage{amsmath}
\usepackage{graphicx}
\usepackage[colorlinks=true, allcolors=blue]{hyperref}

\title{Excel Spreadsheet Analyzer}
\author[1]{Amir Nassereldine}
\author[2]{Patrick Chen}
\author[1]{Jinjun Xiong}
\affil[1]{
University at Buffalo, NY, USA
}

\affil[2]{Palo Alto High School, CA, USA}

\begin{document}
\date{}
\maketitle
\begin{abstract}

Spreadsheets are widely used  in various fields to do large numerical analysis. While several companies have relied on spreadsheets for decades, data scientists are going in the  direction of using scientific programming languages such as python to do their data analysis due to the support, community, and vast amount of libraries. While using python to analyze a company's spreadsheets some information such as the formulas and dependencies of a cell is lost. We propose a tool that  creates an abstract intermediate representation (AIR) of a spreadsheet. This representation facilitates the transfer from spreadsheets into scientific programming languages while preserving inter-dependency information about data. In addition to that, we build a python library on top of our tool  to perform some data analysis in python. 
\end{abstract}

\section{Introduction}

Spreadsheets are a great  simplistic tool with a quick-and-easy way to analyze a dataset. But for the modern era, with big data and more complex analytics and automation, spreadsheets lack some of the functionalities needed to do such automation. This has led to a shift from using spreadsheets to using simple programming languages with large support of libraries and tools. The shift from spreadsheets stems from their shortcomings. For example, on spreadsheets, there is a wide range of errors and it is estimated that 80\% to 90\% of spreadsheets contain significant errors \cite{pryor}. First, there are \textit{accidental errors} that take the form of typological errors that might be unnoticed, deletion errors like deleting some of the entered data or overwriting it, and modification errors occurring when the formulae on the spreadsheet are incorrectly modified \cite{rajalingham}.
Second, we have \textit{reasoning errors} rooting from ill knowledge on adequately using and analyzing the provided functions. Added to that, \textit{qualitative errors} can also be faced, and they are trickier to spot than the previous errors because they do not immediately produce incorrect numeric values but do significantly affect the overall quality and accuracy of the user’s model \cite{rajalingham}. Those types of errors are directly or indirectly from the user’s end. However, apart from those errors, many people worry that their spreadsheets are in fact producing wrong results \cite{pryor}. The specification is missing from spreadsheets. For example, it is impossible to tell whether a spreadsheet is doing the right thing, in other words, whether it is correct unless you know exactly what it is meant to be doing. You need a specification; you can then check whether the specification has been implemented properly and whether the specification is in itself correct. 

Moreover, Spreadsheets have a missing feature which is auditability. It is a risk because when users have entered data that cannot be checked, it is impossible to say whether the answers are correct. Many models are implemented through a series of linked spreadsheets, and it is impossible to tell whether one of these spreadsheets is correct without being able to check the data imported from the other \cite{pryor}. 

Those errors in addition to weak computational processing and the inability to support large amounts of data would probably have a significant influence on a mass-level use of spreadsheets. For those reasons and more, users (especially in fields of data science or finance) are considering other means to do their analysis. Out of those means rises a programming language whose variety of libraries spans a wide range of statistical, scientific, and machine learning use cases. \textit{Python}\cite{python} is a relatively easy programming language with large community support that is being increasingly used in scientific applications and data science. In Python, there exist various advanced and stable numerical libraries (SciPy\cite{scipy}, NumPy\cite{numpy}, matplotlib\cite{matplotlib}…) which helped make it a convenient and accessible tool for a wide scope of audience. Despite that, the integration of python in the world of statistical modeling is slower than it should be. This delay might be explained by the previous lack of a standardized model for statistical use. However, that delay has been sped up by the recent development of many fields including Bayesian statistics (PyMC \cite{pymc}), econometrics (StaM\cite{slam}), and machine learning (sklearn\cite{scikit-learn}). Hence, python is a superior environment for data analysis, statistical computing, and even financing. 

To overcome the shortcomings of spreadsheets, we provide a tool that eases the transition from spreadsheets to different programming languages by creating an Abstract Intermediate Representation(AIR) for any given spreadsheet while preserving the data dependency information. This AIR representation will contain the data stored, the dependencies between data, and the formula expression for any specified cell, and hence no information about the data is lost(especially dependencies and formulas). The Flexibility with the AIR gives us the privilege to extend it to different statistical languages(such as Python and R language \cite{r}) but for this paper, we only  showcase the use of our tool by porting it into a python library and using it on a sample set of spreadsheets. Our python library provides the ability to access, modify and visualize different formulas, write back to the spreadsheet, and allow for tolerance in missing data.

In this paper, we will start discussing the current approaches for transforming spreadsheets  in Section \ref{sec:related_work}. After that, in Section \ref{sec:approach}, we discuss our approach and the construction of the AIR. In section \ref{sec:case_studies}, we portray a case study of the python library we build. Finally, in Section \ref{sec:conclusion}, we conclude.

\section{Related work}\label{sec:related_work}

Different tools were built to facilitate the use of spreadsheets in another language. Apache POI for example fulfills that task for java. This tool is powerful and allows for the modification  of formulas inside java. Our tool takes a generic approach to this issue by creating an intermediate representation that can be later on transferred to other languages. We choose to showcase that in Python because, as stated earlier, it is widely used in the field of data science in addition  to advantages like convenience, and accessibility. 

There exist different projects and libraries that aim to help spreadsheet users to transfer into python and aid in doing data analysis and statistics. Pandas\cite{pandas}
is one of those libraries that has been widely used for loading spreadsheets into python as a form of dataframe. Inspired by R, pandas introduces a DataFrame class applying all the functions in R and even adding more enhancements to it \cite{McKinney}. In implementing statistical models, pandas rids the users of the need for working on preparing their data as in R \cite{McKinney}. Although the library supports the various Excel versions and has an extension for datatypes supporting dates and lambda functions, the library only considers the raw value of each cell. Any formulas or dependency between two or more cells is lost. Our Approach aims on preserving that information while presenting an intermediary representation for the spreadsheet.

We can also find a variety of  tools that are used to support reading from and writing to spreadsheets. xlutils\cite{xlutils}, openpyxl\cite{openpyxl}, and xlsxwriter\cite{xlsxwriter} are samples for such tools in python that allow for reading and writing into spreadsheets. Such tools merely facilitate reading and writing but, they do not provide assistance with analysis.

Other research approaches have tried to introduce python inside excel. That is by allowing users to call python functions inside of a spreadsheet. For example, PyXLL\cite{pyxll} enables writing Excel add-ins in Python instead of VBA. Python functions can be exposed as worksheet functions, macros, menus and ribbon toolbars. Another example that allows users to call python from excel is xlwings\cite{xlwings} which is an open-source library that automates Excel with Python without the need of using VBA.

\begin{figure}

  \centering
  \includegraphics[width=\linewidth/2]{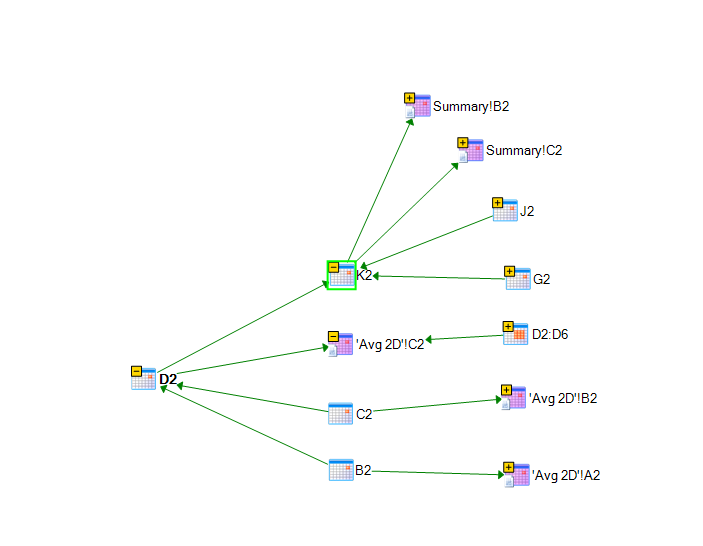}
  \caption{The Inquire Add-in output as result of choosing cell D2 in sheet PXII(see fig.\ref{fig:pxii})}
\label{fig:inquire}
\end{figure}

In addition to those tools, Microsoft Excel\cite{microsoft} provides some add-ons and functionalities that allow for the viewing of formulas in a spreadsheet by pressing ctrl+"`" for example or giving a dependency graph between and across worksheet (see figure \ref{fig:inquire}). However, this is cell/sheet specific in the sense that you have to press on a cell to be able to generate the dependency flow. But as mentioned earlier in our introduction, this permits the possibility of accidental errors.

\section{Excel Analyzer}\label{sec:approach}

\subsection{Abstract Intermediate representation}

In this section, we Explain the different components of the AIR  and the datatypes stored inside of our AIR.
The AIR has two different datatypes: \textit{RawGroups} and \textit{FormulaGroups}.
Both datatypes contain common data fields which are:

\begin{itemize}
    \item \textbf{Name} : The name of the group based on our naming convention
    \item \textbf{Range}: The range of cells contained within the group. We are using a 2D structure for grouping hence the range is of shape Cell1:Cell2 where cell1 is the top left cell of the group and cell2 is the bottom right cell of the group
    \item \textbf{Type}: this contains the type of the cells within the group( for example number, date,...)
\end{itemize}

The difference between the two datatypes is that the first datatype contains cells with raw values that don't include any expressions. While the other contains those that have an expression. The second type of cell is identified by the '=' character which is present at the beginning of the cell as per the spreadsheet convention. Other than that the first datatype consists of an additional data field: 

\begin{itemize}
    \item \textbf{Values }: This is a list of the raw values contained in the group (flattened into 1D)
    
\end{itemize}

While the second datatype contains 3 additional fields:

\begin{itemize}
    \item \textbf{Formula}: This is the normalized form of the function
    \item \textbf{RawFormula}  : This is the raw formula shape of the top left and bottom right raw formulas.
    \item \textbf{Dependencies}:  This is a list of all groups on which this datatype instance depends on (can be both  \textit{RawGroups} and \textit{FormulaGroups})  
    
\end{itemize}

With both these datatypes in mind, our AIR contains a DataFlow Graph which encapsulates the different datatype instances and the relations between them. In this DataFlow Graph, We annotate a node v to be a datatype instance containing the various data fields and the edge e$\{u,v\}$ to represent that node v depends on node u.

Our approach aims to convert any Spreadsheet program into the AIR form automatically. To deal with the complex nature of the instance-based feature of Spreadsheets, a multiple-step approach is proposed to analyze a program systematically. The steps of algorithm \ref{alg:cap} go as follows:

\begin{algorithm}
\caption{The overall algorithm in converting a spreadsheet into AIR}\label{alg:cap}

\begin{algorithmic}[1]

\State\texttt{ $coherentGroups \gets$ \{\};}

\For{ \texttt{every $sheet$ in $spreadsheet$}}

\For{ \texttt{every $cell$ in $sheet$ }}

\State\texttt{ $checked \gets$ 0;}

\If{\texttt{$cell$ starts with '$=$' $\&\&$ $cell$ not checked}}
    \State \texttt{Normalize and Record the cell as an expression cell}

\Else
     \State \texttt{Record the cell as a raw cell}

\EndIf

\EndFor

\EndFor

\For{ \texttt{every $cell$ in $expression cells$ }}
 \State \texttt{Coherence-driven cell grouping}
\EndFor

\For{ \texttt{every $cohenertgroup$ in $coherentGroups$}}
\State \texttt{assign name to group;}
\State \texttt{create node $n$;}
\State \texttt{$dataFlowGraph \gets n$;}

\EndFor

\For{ \texttt{every $node$ in $dataFlowGraph$}}
\State \texttt{create edge based on dependencies;}

\EndFor
\end{algorithmic}
% \todo{Expand graph creation, or make code more abstract}
\end{algorithm}

Our multiple-step approach starts by extracting all the available expressions from a spreadsheet and then normalizing them to a canonical form. After that, we group cells that have coherent logic into one group and find a suitable name for the group. Each group will be a vertex in our AIR dataflow graph and the edges will be the dependencies between the groups. In the following sections, we dive deep into the different components of our approach.

\subsection{Expression Normalizing}

After extracting the expressions (cells that start with '=' character). We  further normalize the expression in two steps. 

In the first step, we will normalize the cells’ locations with respect to the current cell’s location. To do so we use a 3-tuple form  (sheet name, ColumnID, RowID)
For example, assuming the current cell location is A10 in Sheet1, we show this in a tuple as follows: (Sheet1, A, 10). If this cell has a dependency on cell B15 in the same sheet as Sheet1, i.e., (Sheet1, B, 15), this is a relative location reference and we will normalize it in the expression with respect to (Sheet1, A, 10) through a simple
extension of the algebra, i.e., (Sheet1, B, 15) = (Sheet1, A, 10) + (Void, 1, 5).
we use the "\$" to express absolute locations. For example,
if the dependency cell is C\$20 in the same sheet per the expression, its tuple
representation will be (Sheet1, C, \$20), which is an absolute location reference.
In that case, we will normalize it in the expression with respect to (Sheet1, A, 10)
through the same algebra extension while keeping the absolute reference, i.e.,
(Sheet1,C, \$20) = (Sheet1, A, 10) + (Void, 2, \$20). Similarly, if its dependency
cell is D5 but in a different sheet, say Sheet2 (which would have been referred
by the spreadsheet as Sheet2!D5), we will normalize it as $(Sheet2,D, 5) =
(Sheet1, A, 10) + (\$Sheet2, 3,-5)$. In other words, we will keep all the absolute
references while replacing the relative references with algebraic computations
with respect to the current cell’s ColumnID and RowID. We call the second
part of the cell’s location equation as “normalized location expression”.

In the second step, we will normalize the formula expression itself by assigning
a unique canonical variable to a unique dependency cell. For example, if the
formula expression for the current cell A10 in Sheet1 is as follows:$$A10 = Average(B10:B20) / Sheet2!D10+C\$2$$
There are 4 unique cells presented inside of the function  for A10, and
we will assign a unique canonical variable for each, say var0, var1, var2, and var3,
respectively. We would obtain the “normalized formula expression” for A10 as
$$Average(var0:var1)/var2+var3$$
The corresponding association between the canonical variable and its tuple
the representation will be recorded as well.

\begin{figure}
% \clearpage 
  \centering
   \includegraphics[width=\linewidth,trim={0 6cm 0 2cm},clip]{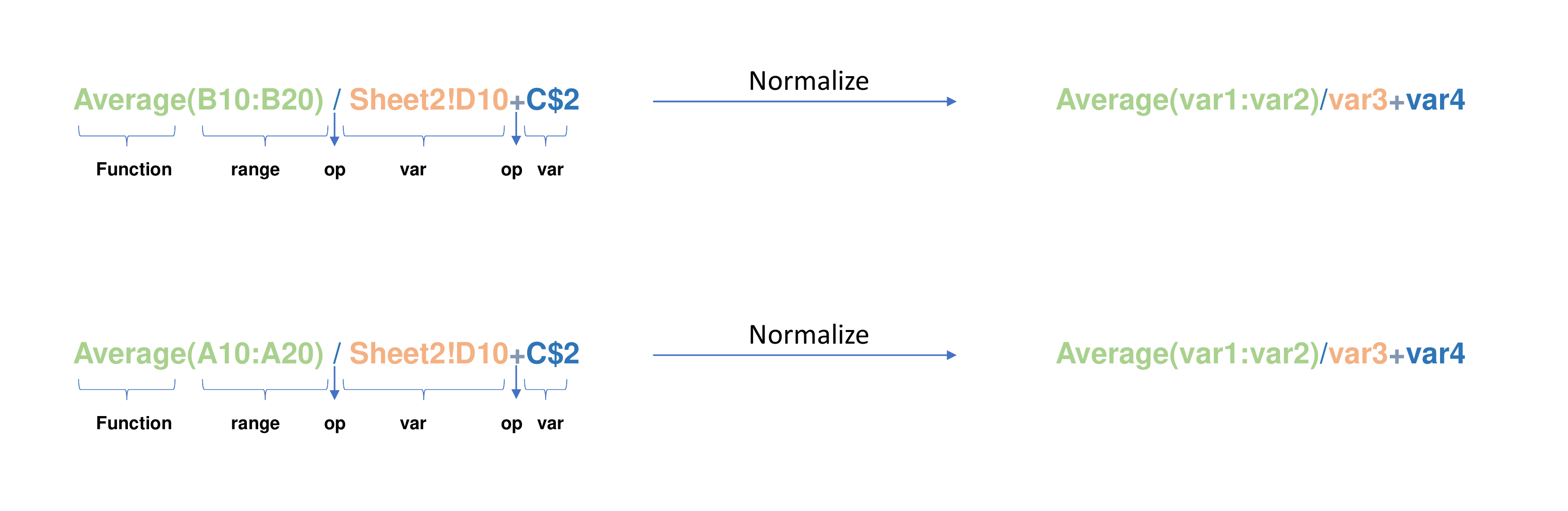}
  \caption{Two coherent formulas}
\label{fig:normalize}
\end{figure}

The purpose of this step is to extract the expression and produce the instance value of any individual cell having its reference location within the spreadsheet. Using the canonical form we are able to find cells containing similar formulas. Figure \ref{fig:normalize} shows an example of 2 similar formulas that have same canonical form. We call the cells of such a case  \textit{coherent cells}.

\subsection{Coherence-driven Cell Grouping}

Another crucial part of our approach is grouping coherent and adjacent cells. 
A constant processing programming logic is applied repetitively across cells in a Spreadsheet because it is based on instance-based data processing. This creates an issue while converting it to a data-flow graph if a naive approach was used because several vertices will have one common computation logic. Therefore, it would be practical to leverage the unique two-dimensional processing characteristics of Spreadsheets. The objective is to detect a bunch of the contiguous regions in a spreadsheet on which similar data transformation would classically reside. Those cells should be logically coherent i.e., they should have equivalent normalized expressions with respect to both normalized formula and location expressions. Such contiguous regions are to be called groups.

Our approach automatically identifies those groups and that they occur in big sizes. “Coherence-driven cell grouping” is what this process is called. This process is done after the normalization of an expression. And for each of the normalized expressions, we search for coherent cells in two directions (up and right). If only of those directions is coherent with the cell we keep adding the cells of that direction to the group until the coherence is broken. But we allow for a tolerance of missing data (explained in section \ref{sec:tolrance}). In case both directions have adjacent coherent-cell coherent we set the bottom right cell to be the intersection of the right cell's column and the lower cell's row. And check if all cells between the current cell and the bottom right cell are coherent.

With the logically coherent cells being grouped, there are a few more steps to
be explored in order to build a fully functional AIR for Spreadsheet.
The first one consists of knowing how to compactly represent all the logically coherent cells in each group and figure out their dependent group (i.e., cells in those groups are referred to by the current group’s cells). This will help us to find a pair-wise
dependency between groups.
The second one is  building a complete data-flow graph from the pair-wise dependency information between two groups. The solution is not that
straightforward as we also need to resolve the dependency conflict among groups.
For example, it is plausible that two groups may depend on the same
group, but different portions of the cells within the group. There is a need
for us to find those conflicts efficiently and resolve the conflicts by splitting the
groups as needed.
The third step is to assign meaningful variable names to each group (or
vertex) in the data-flow graph, and convert the so-obtained data-flow graph to
the desired AIR form.This is later explained in section \ref{sec:naming}

\subsection{Naming Convention}\label{sec:naming}
For each group, we try to provide a name based on our naming convention to be able to access it easily. The first part of the group name is always the sheet name.
Then for any  abstracted data group, if it is 1D(a single row or column), we will search for a name for an adjacent cell according to the direction. Meaning for rows we check to the left of the leftmost upper cell in the group for a string while for columns we check the upper cell the cell above the leftmost upper cell in the group. For example, in \ref{fig:pxii} the cells of column D will be grouped into one group with the name \textit{PXII.PIO }
If a string is not found we the corresponding cells we use the concatenation of the string "Col\_" with ColumnID or the string "Row\_" and the RowID for columns and rows respectively.
For example, in \ref{fig:pxii} if the was NO "PIO" string in cell D1 the cells of column D will be grouped into one group with the name  \textit{PXII.Col\_D}

If the group is 2D we check 2 cells for strings: The cell above the leftmost upper cell in the group and the cell to the left of the leftmost upper cell in the group. The highest priority is for the upper cell then the left cell. If there is a string we use that to name our group. 

If none of the above scenarios apply  we use a global name generator with incremental numbering.

\subsection{Tolerance of Missing Data}\label{sec:tolrance}
An extra feature our tool provides is the ability to provide a threshold for nans or missing data. otherwise, a whole column with the same canonical formula may be divided into several groups because of an empty cell. To do that, we expand our find coherent-driven cell grouping to be able to neglect as many empty cells as provided by a threshold when creating the AIR. Note that the threshold here is only for empty cells.
For example, in Clipper Sheet (see fig.\ref{fig:clipper}) rows 16 and 17 are missing. Having no tolerance will result in having the 3 different column formula groups be divided into 2 groups each. But if the threshold is provided to be a number greater than 2 then we will have a total of 3 forumla groups only.

\section{Case Studies}\label{sec:case_studies}

\begin{figure}[!ht]
% \clearpage 
  \centering
  \includegraphics[width=\linewidth]{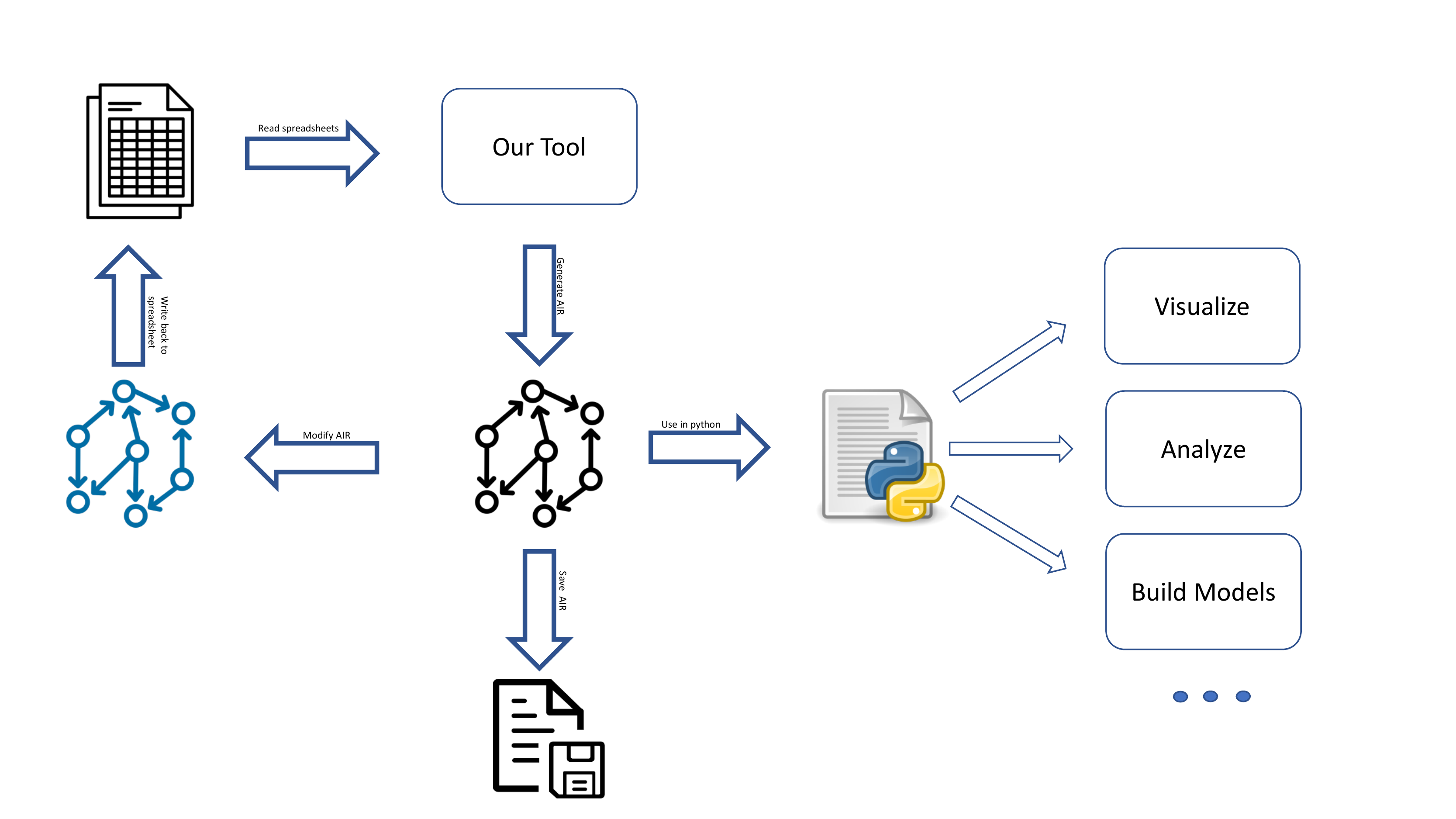}
 
  \caption{ Given a spreadsheet program it reads the spreadsheet and generates AIR based on algorithm  \ref{alg:cap}. After that, we can use the tool in python for various activities such as analyzing, visualizing, and modifying the AIR. 
   \label{fig:flow}
If any modification is done to AIR we are able to save it or port it back to the spreadsheet form.}

\end{figure}

To showcase the value of our approach we build a tool that helps in migrating to python, and in this section, we show  a sample exercise of the tool in python. The flow of our tool is present in figure \ref{fig:flow}

In our exercise, we are given a file \textit{sample.xlsx} which contains 4 spreadsheets: PXII (fig. \ref{fig:pxii} ), Clipper (fig. \ref{fig:clipper}), flash (fig.\ref{fig:flash}), and summary (fig.\ref{fig:summary}). Each of those spreadsheets has inner  dependencies while the summary also depends on all the other 3 spreadsheets. The spreadsheets are provided in appendix A.

We first start by loading the data. Given the name of file and a threshold (2 in this case) the AIR is automatically constructed. 

\begin{lstlisting}[language=Python]
from spreadsheet_graph import SpreadsheetGraph

#Initializing an AIR instance from a spreadsheet
g = SpreadsheetGraph(filename="sample.xlsx",thershold=2)

g.print_graph()
\end{lstlisting}

The output of the above code sample loads the graph with threshold 2 and  outputs the following(the output is cropped for convenience):

\begin{lstlisting}[language=Python]
FORMULA GROUP: (PXII.Pdd PXII!D2:D30)
	RAW GROUP: (PXII.Idd PXII!B2:B30)
	RAW GROUP: (PXII.Vdd PXII!C2:C30)
FORMULA GROUP: (PXII.PIO PXII!G2:G30)
	RAW GROUP: (PXII.IIO PXII!E2:E30)
	RAW GROUP: (PXII.VIO PXII!F2:F30)
FORMULA GROUP: (PXII.PM PXII!J2:J30)
	RAW GROUP: (PXII.IM PXII!H2:H30)
	RAW GROUP: (PXII.VM PXII!I2:I30)
                ...
RAW GROUP: (Flash.VIO Flash!C2:C30)
RAW GROUP: (Flash.IM Flash!E2:E30)
RAW GROUP: (Flash.VM Flash!F2:F30)

\end{lstlisting}

For every raw, and formula group it shows the name and range separated by a space. In addition, for every formula  group, it shows the dependencies on other groups. For example, PXII.Pdd( as explained before PXII is name of sheet while Pdd is the groups name) which is a formula group ranging from D2 to D3 depending on the two raw groups PXII.Idd and PXII.Vdd.

\begin{figure}[!h]
% \clearpage 
  \centering
  \includegraphics[width=\linewidth]{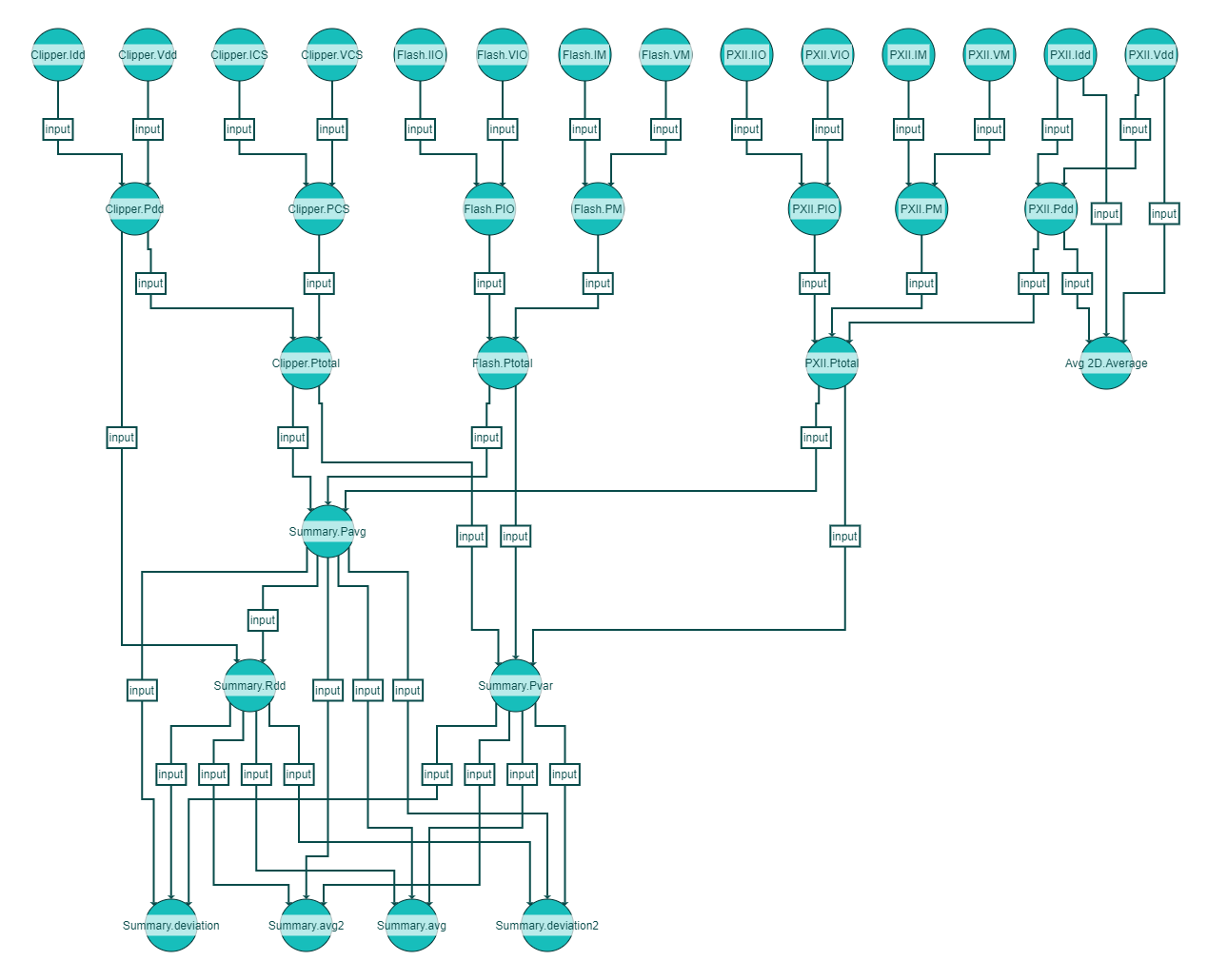}
  \caption{Our flow}
\label{fig:graph}
\end{figure}

To visualize our AIR's dataflow graph we use a python library yfiles\_jupyter\_graphs \cite{yworks}inside a notebook environment\cite{jupyter}  and run the following command.

\begin{lstlisting}[language=Python]
g.get_graph_widget()
\end{lstlisting}

and the output is seen the figure \ref{fig:graph}. Notice how the groups Clippers.Pdd, Clipper.PCS and Clipper.Ptotal are not divided.

\begin{figure}[t!]
    \centering
    \begin{subfigure}[b]{0.4\textwidth}
        \centering
        \includegraphics[width=\textwidth]{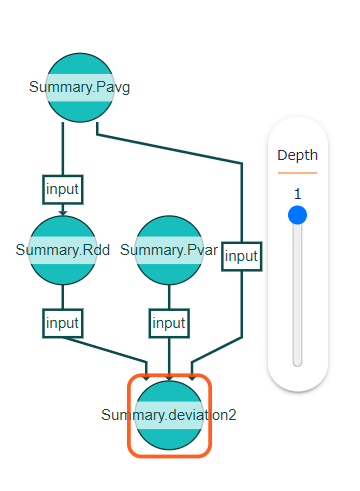}
        \caption{Summary.deviation2 neighbors}
    \end{subfigure}%
  \hfill
    \begin{subfigure}[b]{0.4\textwidth}
        \centering
        \includegraphics[width=\textwidth]{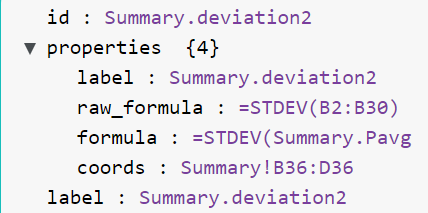}
        \caption{Summary.deviation2  data}
    \end{subfigure}
    \caption{Summary.deviation2 visualization}
    \label{fig:sum}
\end{figure}

Using the same tool we can zooming into one of the nodes (summary.deviation2) and see the neighbors of that node and the data contained inside of it.(check figure \ref{fig:sum})

\clearpage
As part of our design and naming convention we allow the use the formula names to access we use maps(dictionaries)

\begin{lstlisting}[language=Python]
g['Clipper.Ptotal']
#output: FORMULA GROUP: (Clipper.Ptotal Clipper!H2:H30)

g['Clipper.Vdd']
#output: RAW GROUP: (Clipper.Vdd Clipper!C2:C30)

g['Clipper.Ptotal'].formula
#output:'=SUM(Clipper.Pdd,Clipper.PCS)'
\end{lstlisting}

In addition to that, we  also facilitate the search of coherent group given a cell location.
\begin{lstlisting}[language=Python]
print(g.find_group(sheet="Flash", coord="B30"))
#output: RAW GROUP: (Flash.IIO Flash!B2:B30) 
\end{lstlisting}

Another fucntionality in our tool is to be able to modify of formulas as presented  in code sample bellow
\begin{lstlisting}[language=Python]
g["Clipper.Ptotal"].formula = "=(Clipper.PCS + PXII.Pdd)*2 * Flash.PM" 
g.add_group(sheet="Summary", header = "Rdd2", range = "E2:E30", \ 
formula="=cos(PXII.Ptotal * Flash.Ptotal)")
g["Summary.avg"].formula = "=average(Summary.Pavg[1:28])"
g["Clipper.Pdd"].formula = "=Clipper.Idd * Clipper.Vdd * 2"
#Rewrite the cells
g.rewrite_cells()
\end{lstlisting}

 we can finally save our changes back to excel and dump the AIR into a file using the following code
\begin{lstlisting}[language=Python]
g.save(file="sample_Edited.xlsx")
\end{lstlisting}

\section{Conclusion}\label{sec:conclusion}

To solve the problem of losing information when migrating from a spreadsheet to different data analysis means, we presented a tool that generates an abstract intermediate representation of a spreadsheet while preserving the dependency information. We explained the structure of our AIR along with the steps involved in creating such a representation. These steps start by extracting and normalizing cells, then  finding coherent and adjacent cells, followed by creating a name for each group cell, and finally building a dataflow graph. Our tool's AIR takes into consideration having a tolerance for missing data. On top of our tool, we build a python library that shows the application of this tool in python. 

\clearpage
\bibliographystyle{alpha}
\bibliography{sample}
\clearpage

\appendix

\section{Appendix A}
\label{sec:appendixa}
\begin{figure}[ht]
% \clearpage 
  \centering
  \includegraphics[width=\linewidth]{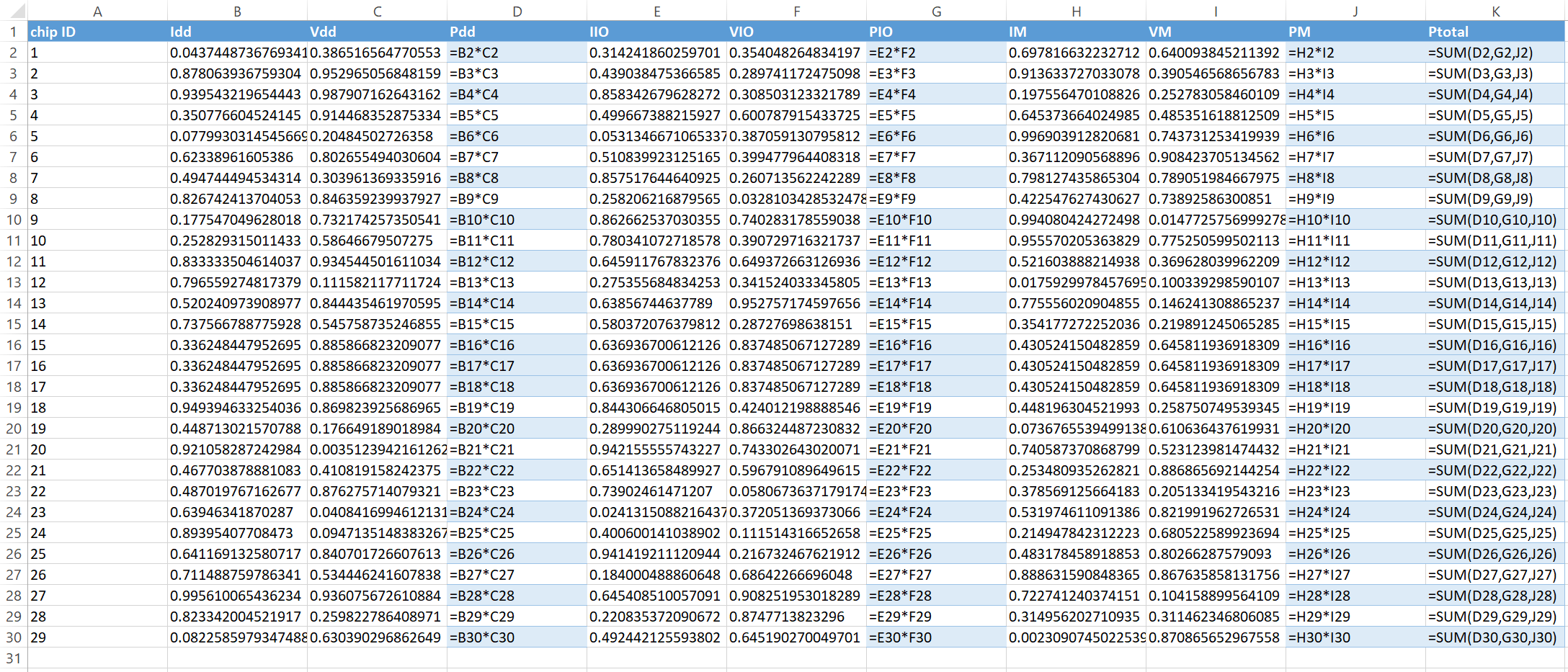}
  \caption{PXII sheet}
  \label{fig:pxii}
\label{sec:appendix}
\end{figure}

\begin{figure}[h]
% \clearpage 
  \centering
  \includegraphics[width=\linewidth]{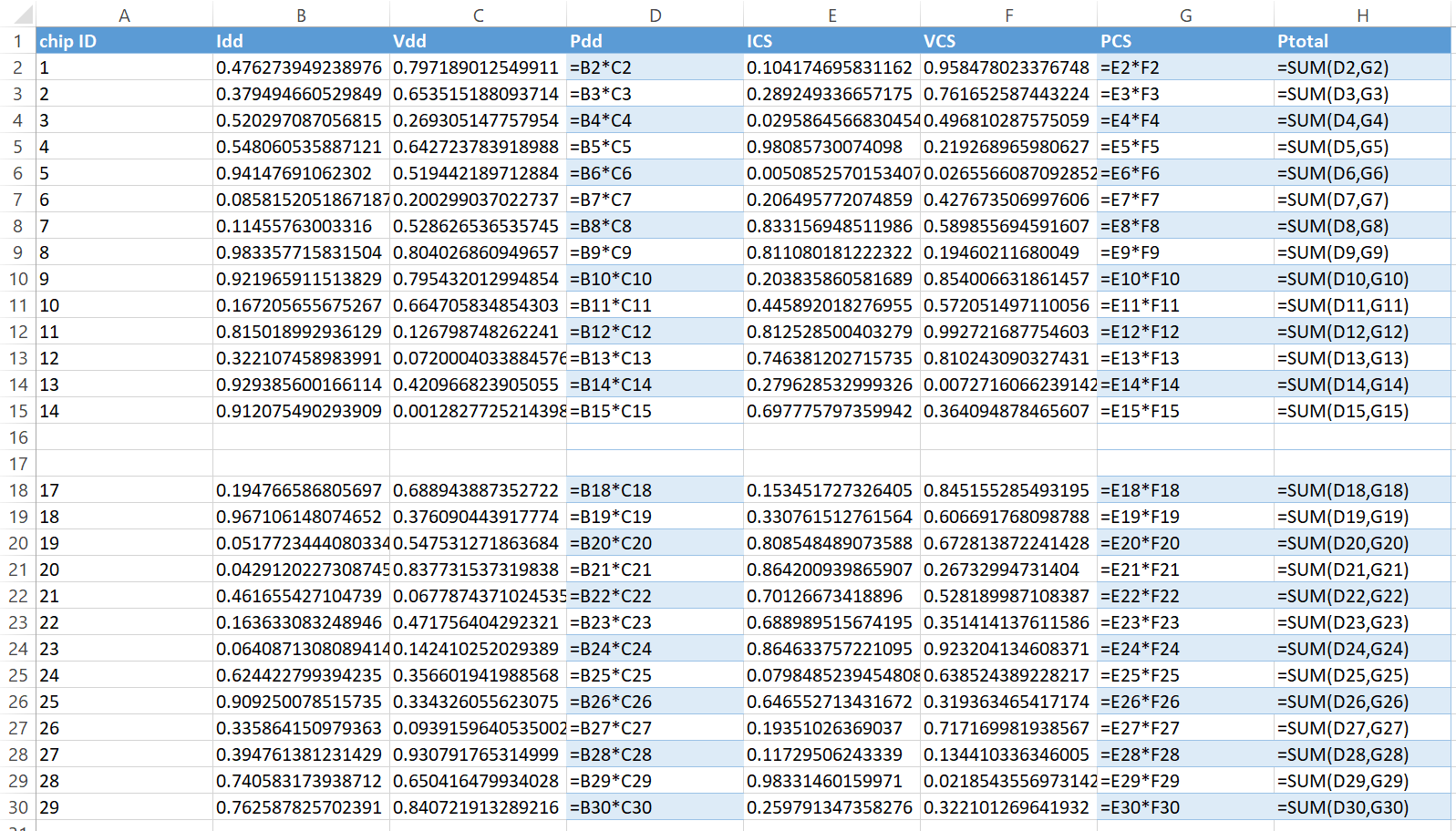}
  \caption{Clipper sheet}
\label{fig:clipper}
\end{figure}

\begin{figure}[h]
% \clearpage 
  \centering
  \includegraphics[width=\linewidth]{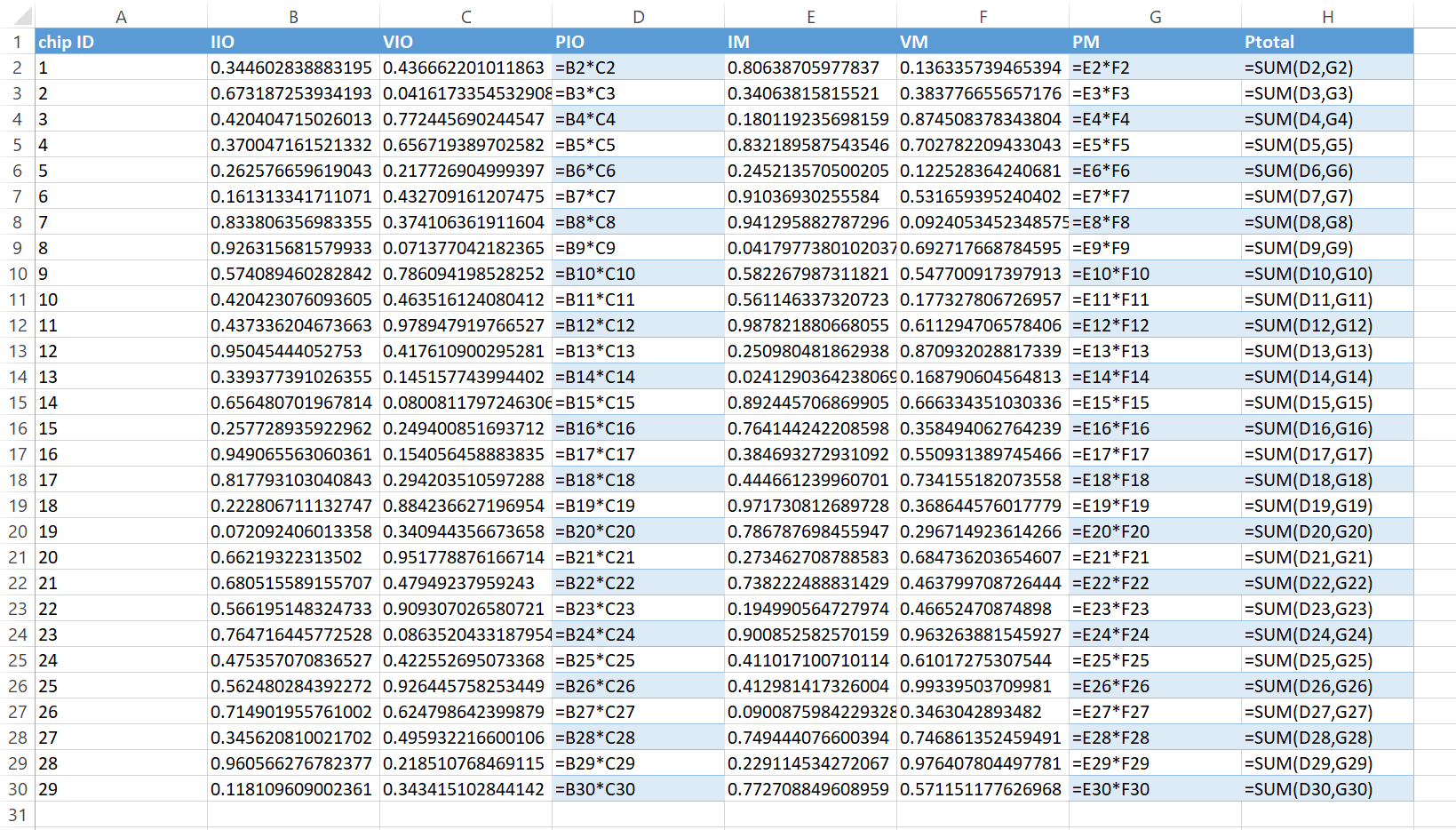}
  \caption{Flash sheet}
\label{fig:flash}
\end{figure}

\begin{figure}[h]
% \clearpage 
  \centering
  \includegraphics[width=\linewidth]{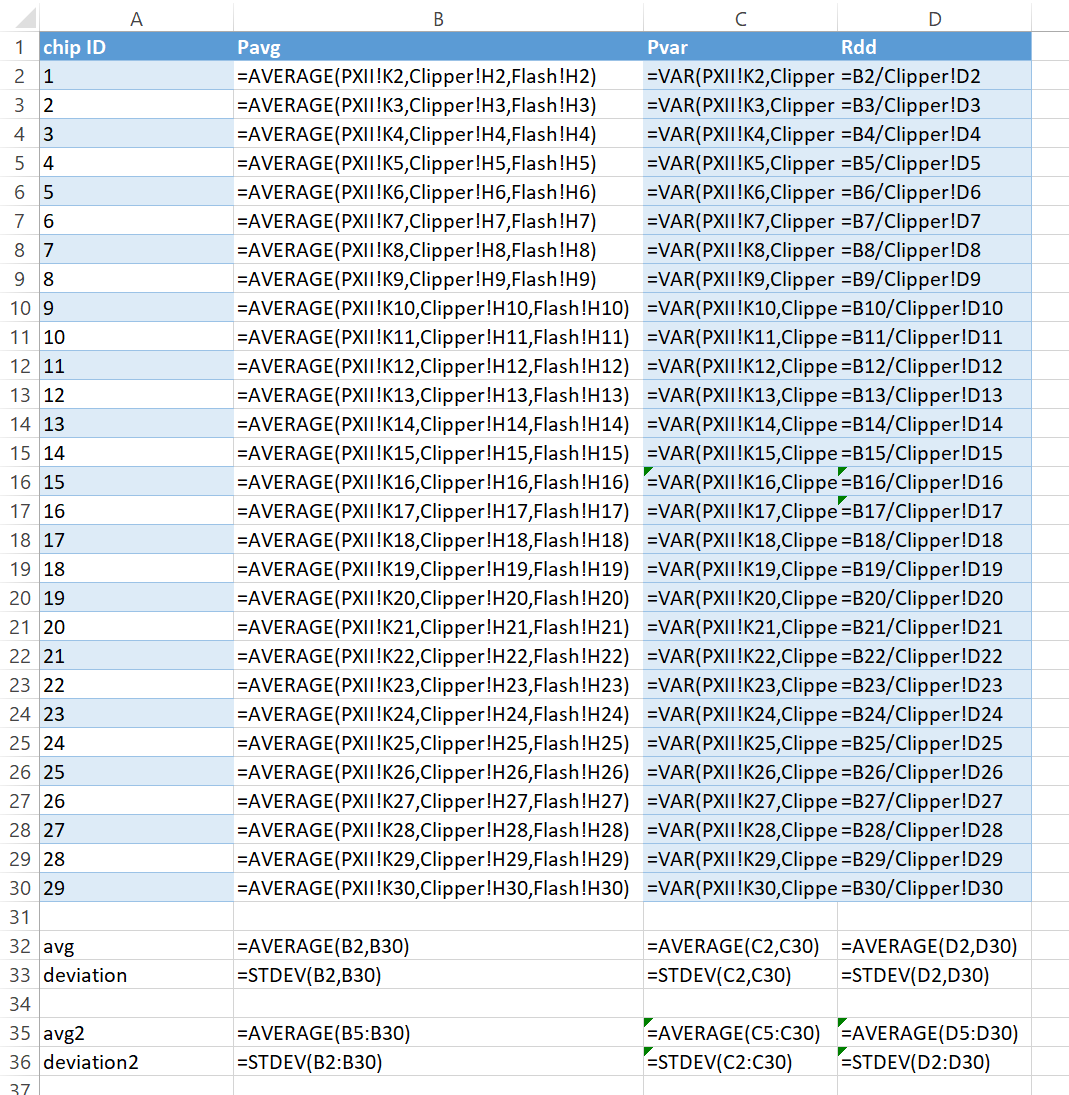}
  \caption{Summary sheet}
  \label{fig:summary}
  \end{figure}

\end{document}